\documentclass[epjST]{svjour}
\usepackage{graphics}
\usepackage{amsmath}
\usepackage{comment}
\usepackage{amssymb}
\begin{document}
\title{Nonlocal-coupling-based control of stochastic resonance}

\author{Vladimir V. Semenov\thanks{\email{semenov.v.v.ssu@gmail.com}}}
\institute{Institute of Physics, Saratov State University, Astrakhanskaya str. 83, 410012 Saratov, Russia}
\abstract{
It is shown that nonlocal coupling provides for controlling the collective noise-induced dynamics  in the regime of stochastic resonance. This effect is demonstrated by means of numerical simulation on an example of coupled overdamped bistable oscillators. In particular, it has been established that increasing the coupling radius and coupling strength allows to enhance or to suppress the effect of stochastic resonance which is reflected in the evolution of the dependence of the signal-to-noise ratio (SNR) on the noise intensity for varying coupling parameters. Nonlocal coupling is considered as an intermediate option between local and global (pairwise or higher-order interactions) coupling topologies which are also discussed in the context of the stochastic resonance control.
} 
\maketitle

Stochasticity represents an intrinsic peculiarity of synergetic systems inevitable affecting the processes of self-organization \cite{haken1983,mikhailov1996}. Moreover, the action of internal fluctuations and external random perturbations can cause a dynamical system to exhibit nontrivial phenomena, including stochastic resonance \cite{gammaitoni1998,anishchenko1999} manifested by the increasing regularity of a non-autonomous stochastic system response to an input signal for increasing noise intensity. Since the first reports in the early eighties \cite{nicolis1981,nicolis1982,benzi1982}, stochastic resonance has become an interdisciplinary fundamental phenomenon uniting a broad variety of stochastic processes of different nature. A spectrum of the stochastic resonance manifestations includes the periodicity of the ice ages \cite{nicolis1981,nicolis1982,benzi1982}, stochastic response of forced lasers \cite{mcnamara1988,fioretti1993}, electronic circuits \cite{luchinsky1999-1,luchinsky1999-2,calvo2006} and neural cells \cite{haenggi2002}, effects in nonlinear chemical reactions \cite{leonard1994,foerster1996,guderian1996,hohmann1996,yang1999}, mammalian auditory system functioning \cite{hong2006}, the anomalous warming of equatorial waters of the Pacific \cite{stone1998} and dozens of other examples.   

Initially revealed and investigated in the context of single oscillators, stochastic resonance is also observed in bistable media \cite{benzi1985,jung1995,wio1996,wio2002,lai2009}, ensembles and networks of coupled oscillators of different topology \cite{gosak2011,tang2012,semenov2022} and time-delay oscillators considered as a spatially extended system by means of virtual space–time representation \cite{semenov2016,semenov2025}. This effect can be exhibited in itself as well as accompany effects of pattern formation, such as spiral wave excitation \cite{jung1995} and chimera states \cite{semenov2016}.

Scientific significance of the issues addressing the stochastic resonance control is dictated by a wide spectrum of practical applications including mechanical energy harvesting \cite{zheng2014}, acoustic weak signal detection \cite{shuyao2016}, machinery fault detection \cite{qiao2019}, spatial pattern writing for semiconductor and optoelectronic device fabrication \cite{roy2019}, Josephson-junction-based threshold detection \cite{ladeynov2023}, information transmission by sensory neurons \cite{longtin1991}, just to name a few (see also the list of references in the review by Gammaitoni et al. \cite{gammaitoni1998}). Characteristics of noise-induced oscillations in the regime of stochastic resonance can be controlled by using different approaches which makes be possible to suppress and to enhance the effect. For this purpose, one can apply time-delayed feedback \cite{mei2009,jia2010}, vary the statistical characteristics of noise, such as the correlation time \cite{haenggi1993} or the probability density function \cite{dybiec2006,dybiec2009}, simultaneously apply additive and multiplicative noise sources \cite{qiao2016}. In the case of coupled oscillators, one can adjust the coupling strength or use the intrinsic properties of the coupling topology \cite{neiman1995,nicolis2017,semenov2022}.

In the current paper, the stochastic resonance control is considered in the context of the coupling topology. In particular, the intrinsic properties of nonlocal coupling are discussed. Such kind of interaction is known to affect the spatio-temporal dynamics of bistable media associated with propagating fronts \cite{colet2014,gelens2014,siebert2014,siebert2015} and stochastic resonance \cite{castelpoggi1998}. However, effects in bistable media with nonlocal spatial interaction can differ from ones exhibited by ensembles of nonlocally coupled bistable oscillators which was recently demonstrated on an example of deterministic and stochastic wavefront propagation \cite{semenov2024}. The second kind of the coupling topology examined in the present paper involves collective connections at the level of groups of nodes called higher-order interactions \cite{battiston2020,boccaletti2023,battiston2022}. Well-studied in the context of synchronization of regular self-oscillations, the impact of higher-order interactions has not been explored in terms of stochastic resonance in ensembles of bistable oscillators. Based on the study of different options for coupling and comparative analysis of the obtained results with the materials corresponding to local \cite{semenov2022} and global pairwise interactions (for instance, see Refs.\cite{jung1992,pototsky2008}), generalized conclusions on the influence of the coupling topology on the stochastic resonance are formulated, which is the main goal of the current paper.

\section{Model and methods}
\label{model_and_methods}
In the present paper, systems under study are ensembles of overdamped bistable oscillators exhibiting the coexistence of two stable steady states. In particular, each oscillator represents the Kramers oscillator including an additive source of noise which is a classical example of the stochastic bistable system describing Brownian motion in a double-well potential \cite{hanggi1990,kramers1940,freund2003,pankratov1999}. All the oscillators are driven by a common external periodic force. In a general form, system equations take the following form:
\begin{equation}
\label{eq:general}
\dfrac{dx_{i}}{dt}=mx_{i}-x_{i}^3+A\sin(\omega_{\text{e}} t)
+\sqrt{2D}n_i(t)+f_i(x_1,x_2,...,x_N), 
\end{equation}
where $x_i$ are the dynamical variables, $i=1, 2, ..., N$ ($N$ is the total number of oscillators),  $f(x_1,x_2,...,x_N)$ is the coupling term. In this study, all the networks consist of $N=100$ elements. Parameter $m$ determines the dynamics of an individual network element. It defines whether the individual element is monostable ($m<0$) or bistable ($m>0$). In the current study, all the elements are in the bistable regime, $m=0.25$, which corresponds to the coexistence of two stable steady states $x^*_{1,2}=\pm0.5$ in the phase space of a single oscillator. Further, terms $\sqrt{2D}n_i(t)\in\mathbb{R}$ describe Gaussian white noise with intensity $D$, i.e., $<n_i(t)>=0$ and $<n_i(t)n_{j}(t)>=\delta_{ij}\delta(t-t')$, $\forall i,j$. The external periodic forcing amplitude is fixed, $A=0.04$, such that the amplitude is less than the threshold value inducing transitions between the states $x^*_{1,2}$ in the ensembles\footnote{It is important to note that the effect of stochastic resonance also can be observed when the driving amplitude is significantly larger than the  threshold value \cite{pankratov2002}.}. The external forcing frequency is assumed to be low, which allows to make the stochastic resonance more pronounced, $\omega_{\text{e}}=0.005$.

The ensembles under study are considered by means of numerical simulations. Numerical modelling is carried out by integration of studied differential equations using the Heun method \cite{mannella2002} with the time step $\Delta t=0.001$ and the total integration time $t_{\text{total}}=10^6$. The used initial conditions are chosen to be random and uniformly distributed in the range  $x_i(t=0)\in[-0.75:0.75]$. 
Results of numerical simulations are visualized as space-time plots of $x_i(t)$ and time realizations of $\overline{x}(t)$ which is an averaged over the ensemble value of dynamical variables $x_i$ at time moment $t$, $\overline{x}(t)=\dfrac{1}{N}\sum\limits_{i=1}^{N}x_i(t)$, and characterises the global instantaneous state of the ensemble. 

Similarly to paper \cite{semenov2022}, the power spectrum averaged over the ensemble, $\overline{S}(\omega)$, is taken into consideration to emphasize the similarity of the effects observed in ensembles with the stochastic resonance in single oscillators: $\overline{S}(\omega)=\dfrac{1}{N}\sum\limits_{i=1}^N S_i(\omega)$, where $S_i(\omega)$ is the power spectrum of the individual element oscillations $x_i(t)$. Then the power spectrum evolution caused by the noise intensity growth fully corresponds to classical stochastic resonance (see Fig. 1 in Ref. \cite{semenov2022} and Fig.~\ref{fig1} in the current paper).

In addition, the signal-to-noise ratio (SNR) is introduced to quantitatively describe the evolution of the noise-induced dynamics caused by varying the coupling parameters and topology. Here, the SNR is introduced in terms of radiophysics and electronics. The power spectra of each oscillator in the ensembles include the spectral peak $S_{\text{max}}$ at the frequency of the external forcing, which is also the main peak in the power spectra. The power spectra also have a minimum $S_{\text{min}}$ close to the main spectral peak ($S_{\text{max}}$ and $S_{\text{min}}$ are schematically illustrated in Fig.~\ref{fig1}~(e) on the example of the averaged spectrum. Technically, $S_{\text{min}}$ is calculated as a mean value of spectral components $S(\omega)$ in the range [$0.75\omega_{\text{e}}:0.85\omega_{\text{e}}$]$\land$ [$1.15\omega_{\text{e}}:1.25\omega_{\text{e}}$] which excludes the neighbourhood of the main spectral peak at the external forcing frequency $\omega_{\text{e}}$. One of the most common SNR's definitions is $\text{SNR}=P_{\text{S}}/P_{\text{N}}$, where $P_{\text{S}}$ is the power of the signal and $P_{\text{N}}$ is the noise power. Then the following formula of $\text{SNR}$ describes the regularity of the experimentally acquired harmonic signal: $\text{SNR}=H_{\text{S}}/H_{\text{N}}$, where $H_{\text{S}}$ is the height of the spectral line above the background noise level in the power spectrum, and $H_{\text{N}}$ is the background noise level close to the resonance frequency $\omega_{\text{e}}$. Thus, in terms of power spectra, the formula for the $\text{SNR}$ takes the form $\text{SNR}=(S_{\text{max}}-S_{\text{min}})/S_{\text{min}}$. The consideration of the $\text{SNR}$ as a function of noise intensity $D$ allows to obtain a non-monotonic curve being a signature of stochastic resonance: there exists an appropriate noise intensity level corresponding to the maximal $\text{SNR}$. To analyze the collective dynamics in ensembles (\ref{eq:general}), the power spectra for all the oscillators are computed by using time realizations $x_i(t)$ and then the corresponding SNRs are calculated. After that, the mean value of the SNR over the ensemble, $\overline{\text{SNR}}$ is extracted. 

\section{Global coupling}
\begin{figure}[t]
\centering
\resizebox{1.00\columnwidth}{!}{\includegraphics{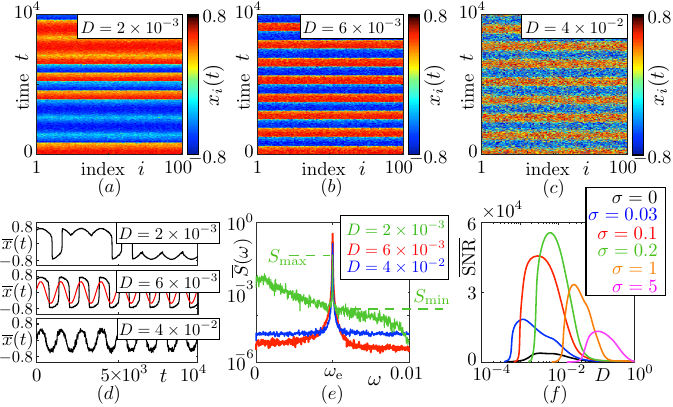} }
\caption{Stochastic resonance in the ensemble of globally coupled bistable oscillators (model (\ref{eq:general}) where the coupling terms are defined by Eqs. (\ref{eq:global_coupling})): (a)-(c) Spatio-temporal dynamics for increasing noise intensity $D$; (d) Time realization of the dynamical variable averaged over the ensemble, $\overline{x}(t)$, for varying noise intensity $D$. The red curve displays the external forcing signal $A\sin(\omega_{\text{e}} t)$; (e) Evolution of the power spectrum averaged over the ensemble, $\overline{S}(\omega)$, for varying noise intensity $D$. System parameters are $m=0.25$, $\sigma=0.2$, $N=100$, $A=0.04$, $\omega_{\text{e}}=0.005$; (f) Dependencies of $\overline{\text{SNR}}(D)$ for varying $\sigma$. Other parameters are the same as in panels (a)-(e).}
\label{fig1}
\end{figure}

Before analyzing the impact of nonlocal coupling topology, model (\ref{eq:general}) is considered in the presence of global coupling. First, it is assumed that the interactions are pairwise:
\begin{equation}
\label{eq:global_coupling}
\begin{array}{l}
f_i(x_1,x_2,...,x_N)= \dfrac{\sigma}{N} \sum\limits_{j=1}^{N}(x_j-x_i),
\end{array}
\end{equation}
where $\sigma$ is the coupling strength. If the coupling strength is large enough and noise is weak, system (\ref{eq:general}) exhibits coherent oscillations characterised by rare or absent transitions between two steady states $x_{1,2}^{*}$ [Fig.~\ref{fig1}~(a)]. Growth of the noise intensity gives rise to more and more frequent transitions between two steady states, while the spatial coherence persists (see the space-time plots in Fig.~\ref{fig1}~(b),(c)). The transitions between two steady states are almost regular for an appropriate noise intensity $D_{\text{opt}}$ [Fig.~\ref{fig1}~(b)] which can vary when tuning the coupling strength. Further increasing the noise intensity decreases the system's response regularity [Fig.~\ref{fig1}~(c)]. The described transformations are reflected in the evolution of the realizations $\overline{x}(t)$ for varying noise intensity $D$. In particular, the realizations $\overline{x}(t)$ undergo transformations which are typical for stochastic resonance in a single bistable system: transitions between basins of attraction of the steady states $x_{1,2}^*$ become more and more frequent, whereas the oscillations are in phase with the external force at an appropriate noise intensity $D_{\text{opt}}$ (see Fig.~\ref{fig1}~(d) where the external signal is depicted in the middle panel corresponding to $D=D_{\text{opt}}=6\times 10^{-3}$).
 
The similarity of the stochastic resonance in the ensemble of globally coupled oscillators and in the single oscillator is complemented by the evolution of the power spectrum $\overline{S}(\omega)$ which fully corresponds to classical stochastic resonance [Fig.~\ref{fig1}~(e)]. First, the height of the spectral peak at the external forcing frequency increases and the peak becomes most pronounced at certain noise intensity $D_{\text{opt}}$. After that, the system’s response to the external periodic forcing becomes less and less regular and one observes inverse transformations of the power spectrum.

The calculations of the mean signal-to-noise ratio involve the averaged power spectrums and the curves $\overline{\text{SNR}}(D)$ obtained at different coupling strengths fully replicate the classical stochastic resonance curves in a single bistable oscillator. As demonstrated in Fig.~\ref{fig1}~(f), increasing the coupling strength allows to make the effect of stochastic resonance more pronounced, which is manifested by the increase of the SNR’s peak values. However, the coupling-induced enhancement of stochastic resonance has a resonant character such that the most pronounced stochastic resonance is observed at $\sigma \approx 0.2$. It must be noted that varying the coupling strength also provides for changing $D_{\text{opt}}$ corresponding to peak values of $\overline{\text{SNR}}(D)$ in a wide range. For instance, $D_{\text{opt}}$ for $\sigma=0.03$ and $\sigma=5$ possesses values $D_{\text{opt}}=10^{-3}$ and $D_{\text{opt}}=10^{-1}$ correspondingly (see Fig.~\ref{fig1}~(f)).

The depicted in Fig.~\ref{fig1}~(f) resonant character of the coupling impact on the stochastic resonance and the ability to enhance it at an appropriate coupling strength are fully dictated by the global character of the coupling topology. Taking into consideration the nonlinear global coupling or modifying it into the second-order interactions does not lead to qualitative changes in the evolution of curves $\overline{\text{SNR}}(D)$ caused by increasing the coupling strength. To visualise this fact, ensemble (\ref{eq:general}) is also studied in the presence of global coupling which involves the second-order interactions: 
\begin{equation}
\label{eq:global_second_order_coupling}
\begin{array}{l}
f_i(x_1,x_2,...,x_N)= \dfrac{\sigma}{2N^2} \sum\limits_{j=1}^{N}\sum\limits_{k=1}^{N}a_{i,j,k}\text{tanh}(x_j+x_k-2x_i),
\end{array}
\end{equation}
where $a_{i,j,k}$ is the adjacency tensor that refers to the 2-simplexes where $a_{i,j,k}=1$ indicates that the node links between oscillators $x_i$, $x_j$, $x_k$ create a triangle. The adjacency tensor is introduced as $a_{i,j,k}=1$ for $i \neq j \neq k$, and $a_{i,j,k}=0$ otherwise to exclude the presence of pairwise interactions which take place in cases $i=j$, $i=k$ and $j=k$.
\begin{figure}[t]
\centering
\resizebox{0.7\columnwidth}{!}{\includegraphics{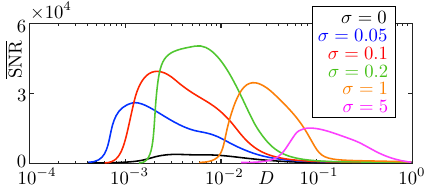} }
\caption{Stochastic resonance in ensemble (\ref{eq:general}) in the presence of second-order interactions (\ref{eq:global_second_order_coupling}) described by the dependencies of $\overline{\text{SNR}}(D)$ for varying coupling strength $\sigma$. System parameters are: $m=0.25$, $N=100$, $A=0.04$, $\omega_{\text{e}}=0.005$.}
\label{fig2}
\end{figure}

Comparative analysis of the dependencies $\overline{\text{SNR}}(D)$ obtained for different strength of pairwise [Fig.~\ref{fig1}~(f)] and second-order [Fig.~\ref{fig2}] interactions allows to conclude that the introduction of the complex nonlinear global coupling organized as interacting oscillator groups does not result in principal changes of the dynamics. Indeed, the introduction of the second-order interactions does not allow to significantly increase or decrease the peak values of the $\overline{\text{SNR}}$ as well as to vary the appropriate noise intensities corresponding to maximal values of the $\overline{\text{SNR}}$.

\section{Local and nonlocal coupling}
\begin{figure}[t]
\centering
\resizebox{0.7\columnwidth}{!}{\includegraphics{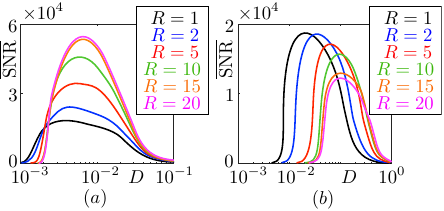} }
\caption{Stochastic resonance in the ensemble of locally and nonlocally coupled bistable oscillators (model (\ref{eq:general}) where the coupling terms are defined by Eqs. (\ref{eq:nonlocal_coupling})): 
dependencies of $\overline{\text{SNR}}(D)$ for varying coupling radius $R$ at $\sigma=0.2$ (panel (a)) and $\sigma=5$ (panel (b)). System parameters are: $m=0.25$, $N=100$, $A=0.04$, $\omega_{\text{e}}=0.005$.}
\label{fig3}
\end{figure}
One can compare the dependencies $\overline{\text{SNR}}(D)$ in Fig.~\ref{fig1}~(f) of the current paper and in Fig.~\ref{fig1}~(g) in Ref. \cite{semenov2022} where the same technique for calculation of $\overline{\text{SNR}}(D)$ was applied for description of stochastic resonance in a ring of locally coupled bistable oscillators (the oscillators' equations and parameter values are the same as in the current paper). The comparative analysis allows to conclude that the global coupling is more effective in the context of the stochastic resonance enhancement as compared to the local interaction. In particular, the maximal peak value of $\overline{\text{SNR}}(D)$ in Fig.~\ref{fig1}~(f) is $\overline{\text{SNR}}(D)=5.3\times 10^4$ (achieved at $\sigma=0.2$), whereas the maximal achieved peak value in the presence of local coupling is $\overline{\text{SNR}}(D)\approx 1.84\times10^4$. However, in case $\sigma>3$ the peak values of $\overline{\text{SNR}}(D)$ achieved at local coupling are higher as compared to the ones obtained in the presence of global coupling. Thus, either local or global coupling topologies are more effective for enhancement of stochastic resonance at different coupling strengths. In such a case, one can expect that increasing the coupling radius as a continuous transition from local to global coupling could enhance or suppress the stochastic resonance which depends on the coupling strength. To analyse this hypothesis, model (\ref{eq:general}) is studied in the presence of nonlocal pairwise coupling: 
\begin{equation}
\label{eq:nonlocal_coupling}
\begin{array}{l}
f_i(x_1,x_2,...,x_N)= \dfrac{\sigma}{2R} \sum\limits_{j=i-R}^{i+R}(x_j-x_i),
\end{array}
\end{equation}
where $R$ and $\sigma$ are the coupling radius and the coupling strength. Character of the stochastic dynamics evolution caused by increasing the coupling radius is illustrated in Fig. \ref{fig3} for two options: $\sigma=0.2$ and $\sigma=5$. In the first case, increasing the coupling radius enhances stochastic resonance [Fig.~\ref{fig3}~(a)]. The effect of saturation is observed at $R>20$ and further growth of the coupling radius does not lead to noticeable changes of the dynamics. 

In contrast, growth of the coupling radius suppresses stochastic resonance at $\sigma=5$ [Fig.~\ref{fig3}~(b)]. Such value of $\sigma$ corresponds to less pronounced global-coupling-induced enhancement of stochastic resonance (compare green and magenta curves in Fig.~\ref{fig1}~(f)), whereas the local coupling provides for much more effective supporting the stochastic oscillation regularity (see Fig.~\ref{fig1} in Ref. \cite{semenov2022}).  As a result, increasing the coupling radius allows to realize continuous transition between local and global coupling and to observe the evolution of the curves $\overline{\text{SNR}}(D)$ such that the stochastic resonance is suppressed, which also has a saturable character: no visible changes of the dynamics are observed when the coupling radius exceeds the value  $R=20$.

\begin{figure}[t]
\centering
\resizebox{1.0\columnwidth}{!}{\includegraphics{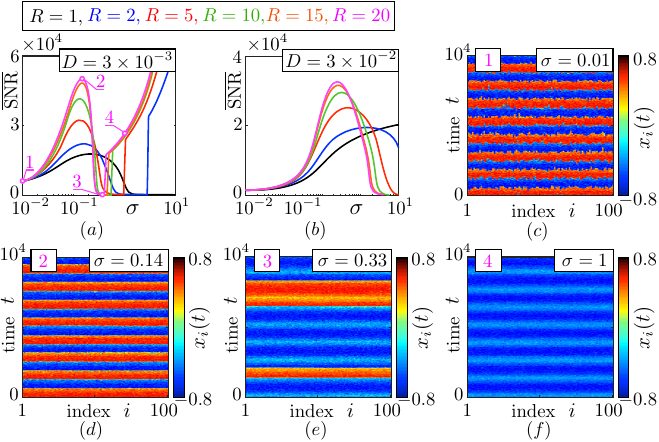} }
\caption{Control of stochastic resonance by varying the nonlocal coupling strength at fixed coupling radius and noise intensity: $D=0.003$ (panel (a)) and $D=0.03$ (panel (b)); (c)-(f) Spatio-temporal dynamics for increasing coupling strength $\sigma$ at fixed $R=20$ and $D=0.003$ (see points 1-4 in panel (a)). System parameters are $m=0.25$, $N=100$, $A=0.04$, $\omega_{\text{e}}=0.005$.}
\label{fig4}
\end{figure}
 
In addition to enhancement and suppression of stochastic resonance, varying the coupling radius and coupling strength allows to shift the values of the noise optimal intensity corresponding to peak values of $\overline{\text{SNR}}(D)$. As a result of such evolution, one more control scheme can be realized. It consists in varying the coupling strength at certain values of the coupling radius and noise intensity. As depicted in Fig.~\ref{fig4}~(a),(b), such approach provides for enhancement and suppression of stochastic resonance. 

In the presence of low noise intensities, one can run into the effect of sharp increasing the signal-to-noise ratios at certain coupling strength level (for instance, see point 4 in Fig.~\ref{fig4}~(a)) after passing the local maximum (enhancement of stochastic resonance, see point 2 in Fig.~\ref{fig4}~(a)) and minimum (suppression of stochastic resonance, see point 3 in Fig.~\ref{fig4}~(a)). This effect is associated with decreasing the frequency of transitions between two basins of attraction (up to extremely low values) with growth of the coupling strength (compare space-time plots in Fig.~\ref{fig4}~(d)-(f)). As a result, oscillations in either potential well are observed in a long time ranges and large values of $\overline{\text{SNR}}$ are not dictated by regular transitions between two steady states $x_{1,2}^*$ and the manifestations of stochastic resonance. This effect can be compensated by increasing the length of considered time realizations $x_i(t)$ (increasing the total integration time) such that one can occasionally observe rare transitions between the steady states and the corresponding $\overline{\text{SNR}}$ becomes significantly lower. However, any real time realization has finite length and one will inevitably deal with single-well dynamics at strong character of interaction.
 
\section{Conclusion}
In the present paper, the effect of stochastic resonance was studied on an examples of coupled overdamped bistable oscillators which is one of the simplest model for demonstration of this phenomenon. Combining the presented results with materials of publications addressing the stochastic resonance in different models (for instance, see Refs. \cite{castelpoggi1998,jung1992}), one can conclude that coupling between oscillators exhibiting the effect of stochastic resonance represents a powerful tool for enhancing the regularity of the noise-induced oscillations. The enhancement can be manifested as increasing the peak values of the signal-to-noise ratio as compared to ones obtained in single oscillators. To achieve maximal values of the signal-to-noise ratio, one can apply different coupling topologies. In particular, global or local coupling can be chosen for this purpose at different coupling strength.

It is important to note that introduction of the higher-order interactions does not give rise to significant changes in the global-coupling-based control of stochastic resonance. In fact, the evolution of the dependencies of the signal-to-noise ratio on the noise intensity caused by growth of the coupling strength in the presence of pairwise and second-order interactions is almost identical. 

Depending on the coupling strength value, either global or local coupling is more effective in the context of the stochastic resonance enhancement at different coupling strength values. Is such a case, transition between two topologies as increasing the radius of the nonlocal interaction provides for enhancing or suppressing the stochastic resonance. This is manifested as increasing or decreasing the peak values of SNR when varying the coupling radius. Similar effect can be achived when the coupling strength varies at fixed coupling radius. Thus, using the nonlocal coupling appears as a promising strategy for controlling stochastic resonance. 

As demonstrated in the current paper, local coupling or low-radius nonlocal interactions are more effective for enhancement of stochastic resonance in the presence of strong interactions (at high values of the coupling strength). This is a promising result in the context of applied science, since the coupling radius can be restricted by technology. For instance, diffusive coupling is typical for Josephson junction arrays \cite{barbara1999,pankratov2017}.

\section*{Declaration of Competing Interest}
The author declares that he has no known competing financial interests or personal relationships that could have appeared to influence the work reported in this paper.

\section*{Acknowledgments}
This work was supported by the Russian Science Foundation (project No. 24-72-00054).

\section*{Data availability statement}
The data that support the findings of this study are available from the author upon reasonable request.

%


\end{document}